\documentclass[twocolumn,superscriptaddress,showpacs,preprintnumbers,amsmath,amssymb,floatfix]{revtex4}
\usepackage[dvips]{graphicx}
\usepackage{amsmath}
\usepackage{bm}
\usepackage[dvips]{graphicx}
\usepackage{amsfonts}
\usepackage{bm}
\begin{document}

\title{Structural arrest in dense star polymer solutions}
\author{G. Foffi}
\affiliation{Dipartimento di Fisica and INFM
Center for Statistical Mechanics and Complexity, 
Universit\`{a} di Roma La Sapienza, Piazzale Aldo Moro 2, I-00185 Rome, Italy}
\author{F. Sciortino}
\affiliation{Dipartimento di Fisica and INFM
Center for Statistical Mechanics and Complexity, 
Universit\`{a} di Roma La Sapienza, Piazzale Aldo Moro 2, I-00185 Rome, Italy}
\author{P. Tartaglia}
\affiliation{Dipartimento di Fisica and INFM
Center for Statistical Mechanics and Complexity, 
Universit\`{a} di Roma La Sapienza, Piazzale Aldo Moro 2, I-00185 Rome, Italy}
\author{E. Zaccarelli}
\affiliation{Dipartimento di Fisica and INFM
Center for Statistical Mechanics and Complexity, 
Universit\`{a} di Roma La Sapienza, Piazzale Aldo Moro 2, I-00185 Rome, Italy}
\author{F. Lo Verso}
\affiliation{Istituto Nazionale di Fisica della Materia
and Dipartimento di Fisica, 
Universit\`a di Milano, Via Celoria 16, I-20133 Milano, Italy}
\author{L. Reatto}
\affiliation{Istituto Nazionale di Fisica della Materia 
and Dipartimento di Fisica, 
Universit\`a di Milano, Via Celoria 16, I-20133 Milano, Italy}
\author{K. A. Dawson}
\affiliation{University College Dublin,
Irish Center for Colloid Science and Biomaterials, 
Department of Chemistry, Belfield, Dublin 4, Ireland}
\author{C. N. Likos}
\affiliation{Institut f\"ur Theoretische Physik II,
Heinrich-Heine-Universit\"at D\"usseldorf, 
D-40225 D\"usseldorf, Germany}

\date{\today}
\pacs{83.60.Hc, 64.70.Pf, 82.70.Dd, 83.80.Uv}

\begin{abstract}
The dynamics of star polymers has been
investigated via  extensive Molecular- and Brownian Dynamics simulations  
for a large range of  functionality $f$ and packing fraction $\eta$.
The  calculated isodiffusivity curves display both
minima and maxima as a function of $\eta$ 
and minima as a function of $f$.  
Simulation results are compared with theoretical 
predictions based on different approximations for the structure factor.
In particular, the ideal glass  transition line predicted by 
mode-coupling theory is shown to exactly track the  
isodiffusivity curves, offering a theoretical understanding 
for the observation of disordered arrested states in
star polymer solutions.

\end{abstract}

\maketitle
%
%
Star polymers play an important role in soft condensed matter systems, since they have been shown to interpolate between hard colloids with a strong repulsive core on one side and the soft flexible polymeric systems on the other \cite{likos01}. 
Star polymers are constituted by a given number of polymeric arms, called functionality $f$, tied to a common center. 
As the functionality increases, the system becomes more similar to a  hard sphere (HS) system, while  lowering the functionality the inter-particle  potential becomes increasingly soft. 
Recently, many interesting properties of star polymers have been clarified 
on the basis of an effective interaction potential 
between star-polymer centers \cite{likos:prl:98}.
In line with their hybrid polymer-hard colloid character,
star polymers display no crystallization transition when the functionality $f$ is low,
$f \leq 34$. 
At higher functionalities, a freezing transition takes place at about the
overlap concentration of the system, into a bcc-solid for
lower functionalities and into an fcc-solid for higher ones \cite{watz:prl:99}.
The freezing is succeeded by either a reentrant
melting transition to the fluid for 
intermediate functionalities, $34 \lesssim f \lesssim 54$,
or by a cascade of structural phase transitions at higher values of $f$. 
The functionality-dependent bcc- and fcc-solids\cite{gast:prl:93,gast:pre:96}, as well as the reentrant  melting transition \cite{gast:macrom:97}, have been 
experimentally observed in solutions of star-like block copolymer micelles.
Though the
crystalline solids are the phases of thermodynamic equilibrium at such high
concentrations, the experimental situation is often somewhat different.
A variety of studies
with star-polymers or star-like systems of 
various functionalities has shown that
it is quite difficult to nucleate a crystal. Especially at high functionalities,
the solutions display a gelation transition, i.e., a dynamical arrest into
an amorphous crowded state in which the characteristic relaxation time of the system becomes extremely long
\cite{vlassopoulos:jpcm:01, kapnistos:prl:00, loppinet:macrom:01, stellbrink:pcps:00, stiakakis:prl:02, stiakakis:pre:02}.
The onset of gelation, as opposed to crystallization, is further enhanced by the presence of some polydispersity in the
samples, which gets indeed more pronounced as functionality increases.

The purpose of this work is to analyze the dynamics
of star polymers in athermal solvents theoretically, by employing a
combination of methods. Using the star-star effective interaction potential, in which
all microscopically fluctuating degrees of freedom are averaged out, we carry
out detailed Molecular Dynamics (MD) and Brownian Dynamics (BD) simulations to
measure the diffusivity of the
star polymer fluids down to the homogeneous nucleation limit. Moreover, we carry
out a Mode-Coupling Theory (MCT) analysis \cite{mct} of the long-time limit of the correlation
functions, which allows us to locate the 
nonergodicity (ideal glass) transition line of the
system.  We find that, on increasing the number density of polymer,
the characteristic time of the system goes through
a sequence of maxima and minima which we show to be
related to the oscillatory behavior of the effective
HS diameter \cite{federica:jpcm:03}. 
We also find a strong correlation between the equilibrium
phase diagrams of the system and the ideal glass line. The 
equilibrium reentrant melting transition  is shown to have
its counterpart in the reentrant melting of the disordered glass
nested between two stable fluid phases.
Finally we discover a striking
similarity in the shape of the isodiffusivity and 
MCT ideal glass curves, regardless of the detailed 
approximation employed in the calculation of the structure factor $S(q)$ 
and the type of dynamics (MD or BD).  
These findings strongly support the interpretation 
that the structural arrest of star polymers 
is a glass transition of  
`effective hard spheres' characterized by a 
$\eta$- and $f$-dependent HS diameter, despite the fact that
the variety and nature of the equilibrium phases
themselves is very different from the HS case.
%
%

Standard MD and BD techniques 
have been employed in order to study the dynamic behavior of the 
star polymer system \cite{md}. 
MD and BD lead to identical predictions for the long-time dynamics of the
system within the framework of the MCT \cite{mct,szamel:loewen:pra:92}.
Thus, though BD offers
a more realistic description of the short-time diffusion of the
particles, the long-time behavior which is the relevant one for
the glass transition is the same in both.
Hydrodynamic interactions are ignored and the analysis is based on
the effective center-center interaction potential 
\begin{eqnarray}
\nonumber
\label{eq-potential}
\beta V(r) & = & {5\over 18}f^{3/2}\left[ -\ln\left({r \over \sigma}\right) + 
{1 \over {1+\sqrt f/2}}\right],~~~~~~~r \le \sigma \nonumber \\
\nonumber
 & = & {{5} \over 18}f^{3/2} {{\sigma / r} \over {1+\sqrt f/2}} 
\exp\left[-{{\sqrt f (r-\sigma)}\over{2\sigma}}\right],~~r \ge \sigma
\\
\end{eqnarray}
where $\beta=1/k_BT$ and  $r$ 
is the distance between the two centers. This  potential is a combination 
of a logarithmic interaction at short distances, 
which gives the interaction its ultra soft character and 
stems from the scaling analysis of Witten and Pincus \cite{witten:pincus:86}, 
and a Yukawa form for the decay at long distances matched 
at a distance $r=\sigma$. 
The quantity $\beta V(r)$ depends 
solely on $\sigma$ and $f$. 
The validity of this potential has been demonstrated via 
extensive comparisons both with small angle neutron scattering 
data \cite{likos:prl:98, stellbrink:pcps:00} 
and with monomer-resolved simulations \cite{arben:macrom:99}. 
In this entropic interaction, $f^{-1}$ plays the role of $T$ in normal fluids.

In order to locate the line of structural arrest, 
the mean squared displacement averaged over all particles
$\langle|{\bf r}_i(t)-{\bf r}_i(0)|^2\rangle$, where ${\bf r}_i$ 
is the position of particle $i$,  
has been computed as a function of time $t$ for several 
values of the two control parameters, 
the functionality $f$ and the packing fraction 
$\eta = \pi \rho \sigma^3/6$, where $\rho$ is the number density.  The values of the self-diffusion coefficient $D$ are
calculated from the relation   $6Dt = \langle|{\bf r}_i(t)-{\bf r}_i(0)|^2\rangle$, valid for large $t$.

\begin{figure}
\begin{center} 
\includegraphics[width=8cm,angle=0.,clip]{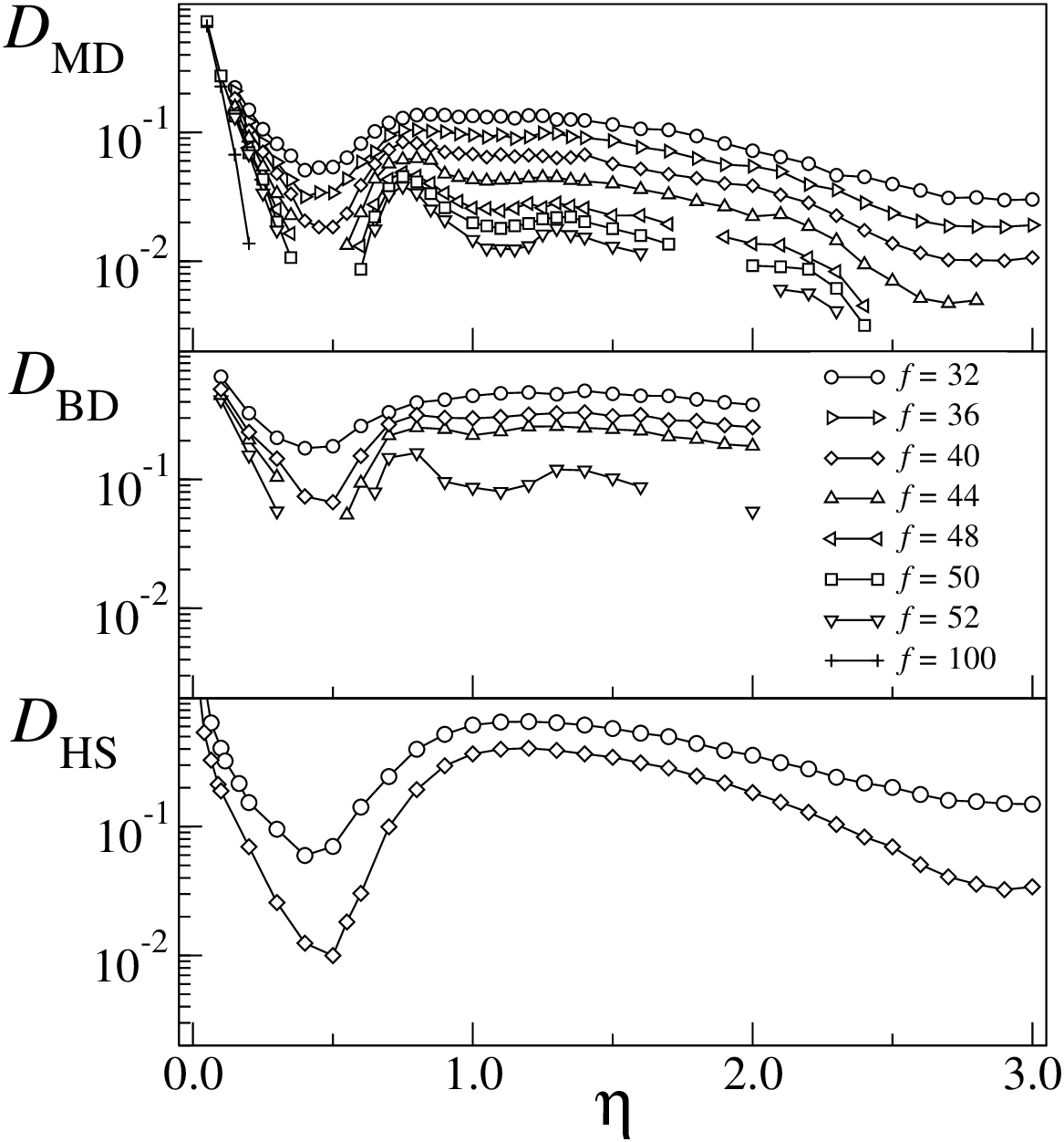}
\caption{
The self diffusion coefficient, $D_{\rm MD}$ obtained by Molecular dynamics 
(upper panel, MD units \cite{md}), 
$D_{\rm BD}$ by Brownian dynamics (middle panel, BD units) 
and $D_{\rm HS}$ for an equivalent HS system 
(lower panel, arbitrary units), as a function of packing 
fraction for various values of the functionality \protect\cite{xtal}. 
Missing points along the simulation curves were identified as crystalline states.
}
\label{fig1:plot}
\end{center}
\end{figure}

%
%
For the same system, the MCT equations \cite{mct} for 
modelling the structural arrest in supercooled liquid states  
have been solved;  MCT furnishes a time evolution equation for the 
normalized density fluctuation correlator $\Phi(q,t)$ as a function 
of the momentum transfer $q$ and $t$, which contains a term 
nonlinear in the correlator itself.  
Knowing $S(q)$ corresponding to the 
given interaction potential, the memory kernel entering 
in the nonlinear term of the MCT equation can be evaluated and the 
equation solved for various values of $q$. 
In particular the non-ergodicity transition leading to 
structural arrest is obtained by performing the limiting value of 
the density correlator 
$\lim_{t \rightarrow \infty} \Phi(q,t) = f(q)$. The non-ergodicity factor $f(q)$  is the solution of the equation
\begin{equation}
\label{eq-nonergodicity}
{{f(q)} \over {1-f(q)}} = m(q)
\end{equation}
where the memory kernel is quadratic in the correlator
\begin{equation}
\label{eq-kernel }
m(q) = {{1}\over{2}}\int {{d^{3}k}\over{(2\pi)^{3}}}{\cal V}( {\bf q},{\bf k})f({k}).
f({|{\bf q}-{\bf k}|})
\end{equation}
The vertex functions ${\cal V}$, the coupling constants of the theory, are
\begin{eqnarray}
\label{eq-vertex}
{\cal V}( {\bf q},{\bf k}) =  \frac{\rho}{ q^{4}}\left[{\bf q}
\cdot( {\bf q} - {\bf k}) 
~c({|{\bf q}-{\bf k}|})+ {\bf q}\cdot {\bf k}~c({k})\right]^{2} \times \nonumber \\
S({q}) S({k}) S({|{\bf q}-{\bf k}|})
\end{eqnarray}
and  depend only on the Fourier transform 
of the direct correlation function $c(q)$, 
or equivalently on $S(q)$.
In the $A_2$ bifurcation scenario of MCT \cite{mct} the 
solutions of Eq.~(\ref{eq-nonergodicity}) jump from zero to a finite value 
at the ideal glass transition. 
The locus of the fluid-glass transition  can be calculated  
varying the control parameters of the system, $f$ and $\eta$. 

%
%
$S(q)$ has been 
calculated for the potential of Eq.~(\ref{eq-potential}) 
using two different approximate closures 
for the Ornstein-Zernike equation.
The first one is the self-consistent Rogers-Young (RY) closure \cite{ry:84},
which satisfies
the thermodynamic requirement that 
the isothermal compressibilities calculated either following
the so-called fluctuation route or the virial route be identical to one another.
The second approach is the modified hypernetted-chain approximation 
(MHNC) \cite{mhnc:79}. In the MHNC, one 
approximates the bridge function of the 
system \cite{hansen:mcdonald} with
that of an effective HS fluid.
The  optimum  value of the diameter  of this reference HS
system is  chosen in such a way to satisfy the Lado criterion \cite{lado:82}.
Both the RY and the MHNC closures have been shown to yield results
that compare extremely well with simulations \cite{federica:jpcm:03, watz:jpcm:98}.

%
%
The diffusion coefficient $D$ for MD and BD  are reported  in Fig.\ \ref{fig1:plot}, in the upper and middle panels respectively for a large set of $f$ and $\eta$ 
values \cite{xtal}.
In simple liquids, $D$ decreases monotonically on increasing $\eta$. In the present system, $D$ has an highly non monotonic behavior. At $f=32$ a minimum around $\eta=0.5$ is followed by a flat maximum up to
$\eta=1.5$.  At larger $f$ values ($f>40$) a clear sequence of minima and maxima is observed.
The MD and BD results show the same trend, and in supercooled states the $D$ values become proportional.
Such a proportionality  supports the interpretation of the present results in term of MCT for supercooled liquids. Indeed, one of the basic prediction of the theory is the independence of the slow dynamics from the microscopic 
dynamics \cite{mct,kobbindernauroth, szamel:loewen:pra:92}. 

\begin{figure}
\begin{center}
\includegraphics[height=6.5cm,angle=0.,clip]{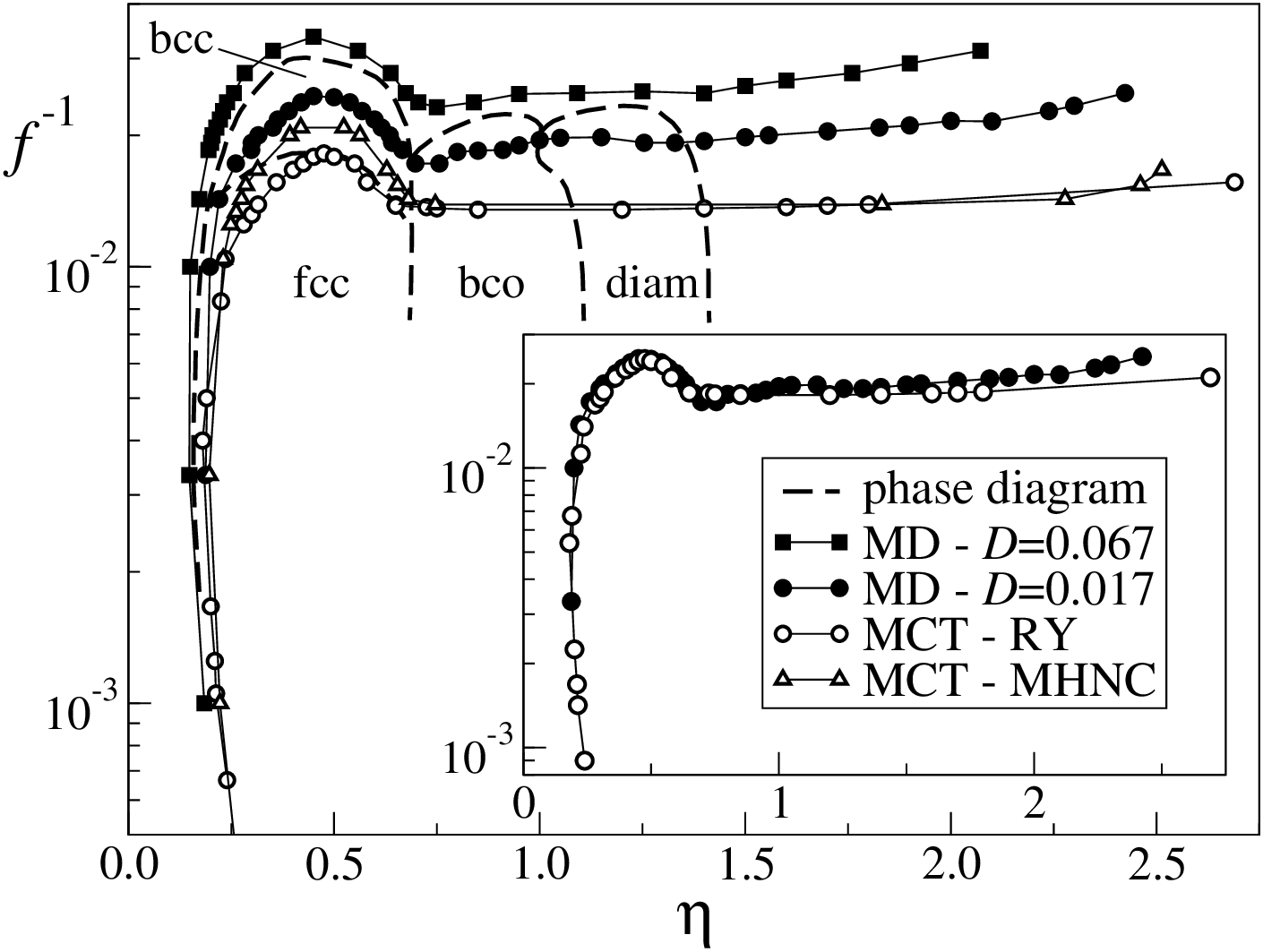}
\caption{
Experimental iso diffusivity curves for two 
values of the MD-diffusion coefficient ($D = 0.067$ and $D = 0.017$) 
and the MCT fluid-glass lines (computed with RY and MHNC).
The equilibrium phase diagram (from Ref.\ \cite{watz:prl:99}) 
is also reported for comparison. 
The inset compares the RY-ideal MCT glass line with the 
isodiffusivity data for $D=0.017$, after a shift along the $f^{-1}$ axis.}
\label{fig2:plot}
\end{center}
\end{figure}

The $\eta$ and $f$ dependence of $D$ can be  rationalized by calculating the isodiffusivity curves \cite{zacca1}, 
i.e. the locus of points in the ($\eta,f$) plane where $D$ has a constant value.
Such curves, for two different values of $D$, are shown in Fig.~\ref{fig2:plot}
together with the equilibrium fluid-crystal coexistence lines,  
as calculated in Ref.\ \cite{watz:prl:99}. The isodiffusivity
curves run parallel to the coexistence lines, suggesting that
crystallization in this system is achieved at the same $D$ value, 
independently from the underlying crystalline phase, and offering thereby
a strong confirmation of the dynamical freezing criterion of 
L{\"o}wen {\it et al.}\ \cite{lowen:prl:93}.
Fig.~\ref{fig2:plot} also shows  the MCT ideal glass transition line 
(which can be considered as the isodiffusivity curve
in the limit $D=0$) using as input  the RY and the  MHNC  $S(q)$.
Both the RY and the MHNC ideal glass lines track the
molecular dynamics isodiffusivity curves. The simulation curve can be perfectly superimposed to the theoretical
curve after a shift in $f^{-1}$, as shown in the inset of
Fig.~\ref{fig2:plot}. The agreement between theory and
simulation below $\eta=1$ is extremely good, independently of the $f$ value. Even the reentrant shape of the isodiffusivity  curve for $f> 500$  is captured by the theory.

The shape of the ideal glass line in the ($\eta,f$) plane
suggests  interesting possibilities for modulating the
dynamics in star polymer systems. If crystallization can be avoided, a glass state can be generated by compression which can then be melted with further compression (for example for $f=36$). Similarly, the dynamics of the polymer can be slowed down  and increased  again upon increase of its
functionality (for example at $\eta=0.23$).

\begin{figure}
\begin{center}
\includegraphics[width=8cm,angle=0.,clip]{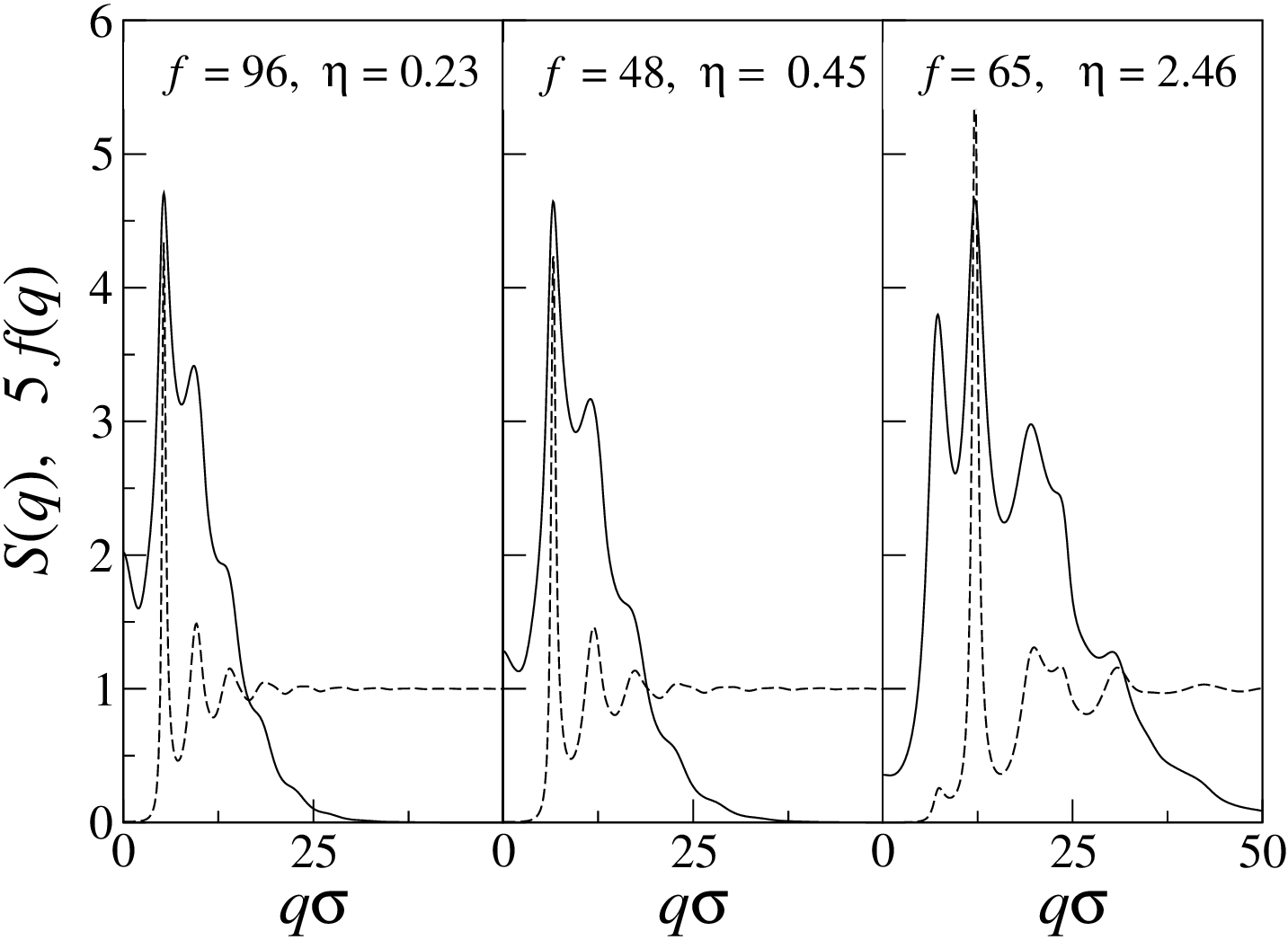}
\caption{
The MHNC structure factor $S(q)$ (dashed line) and the non ergodicity factor $f(q)$ (solid line) for three representative points along the fluid-glass transition line.}
\label{fig3:plot}
\end{center}
\end{figure}

In order to better grasp the peculiar shape of the dynamical arrest curve,  we test the hypothesis that the slowing down of the dynamics  is controlled by the $\eta$ and $f$ dependence of an effective hard core. 
MHNC provides a well defined way for calculating 
the equivalent HS diameter $\sigma_{\rm HS}$ and thus $\eta_{\rm HS}$.
In the case of the star polymer potential, employing the Lado
criterion \cite{lado:82} for the MHNC, 
the dependence of $\eta_{\rm HS}$ on $\eta$ has
been already studied and it has been shown that it reflects the features
of the interparticle interaction \cite{federica:jpcm:03}.
On the other hand, values of the diffusion 
coefficient $D_{\rm HS}$ for HS systems as a function of 
$\eta_{\rm HS}$, the HS packing fraction, 
can be easily obtained directly by MD simulation. 
In a similar way $D_{\rm HS}$ 
is known to follow for BD the simple 
law\cite{pusey98, szamel:loewen:jpcm:93, hinsen:physa:90}
$D_{\rm HS} = D_0 (1-2\eta_{\rm HS})$ for $\eta_{\rm HS} \le 0.4$, with 
$D_0$ a constant.
This offers the possibility of converting the known $\eta$-dependence of the
effective HS packing 
fraction, $\eta_{\rm HS}(\eta)$, into an effective $D_{\rm HS}(\eta)$ of the 
corresponding HS system (both for BD and MD). This, in turn, can
be compared with the star-polymer $D$ values reported here. 
The comparison for MD is shown in the lower panel of 
Fig.\ \ref{fig1:plot}, which illustrates that the simple mapping 
between the packing fractions of the star polymer $\eta$ and the HS 
system $\eta_{\rm HS}$ captures the main features of the diffusion coefficient, 
e.g., the location of the minima and maxima of the curves.
The agreement between the two sets of data  suggests that the 
slow dynamics  in star polymer systems can be traced back 
-- via a density- and functionality-dependent effective HS diameter -- 
to the slow dynamics of the HS system 
(which is accurately described by MCT\cite{goetzepisa}).

Finally, Fig.\ \ref{fig3:plot} shows the 
non-ergodicity factor $f(q)$ at three different 
representative points along the ideal glass line,
calculated with the MHNC $S(q)$.  The $f(q)$ shape changes 
continuously along the ideal-glass line, going from the typical 
HS shape at small $\eta$ (left and central panel) to a much more  
structured shape at large $\eta$
(right panel). The $f(q)$ width is a measure of the
inverse of the cage localization length, which decreases 
on increasing $\eta$.  The $q$ dependence of $f(q)$ is always in phase 
with $S(q)$, a feature common to all previously studied models.  Two 
interesting features which
develop at large $\eta$ are the significant increase in $f(q)$
in the $q$-range at $q \sigma = 7.44$ and the
small value for vanishing $q$.

%
%
The MD and BD simulation data reported in this Letter, 
for a large range of  $f$ and $\eta$, 
show that the dynamics of star polymer solutions is extremely rich.  
The isodiffusivity curves have been shown to display 
both minima and maxima as a function of $\eta$ and minima as a 
function of $f$  which have been successfully connected to 
the behavior to the $\eta$ and $f$-dependence of the 
effective hard core diameter of an equivalent HS system.
The detailed comparison between theoretical predictions 
and simulation confirms that  MCT is a valid approach 
for guiding the interpretation of the disordered arrested 
states of soft matter materials \cite{fsnat,ken}, 
offering a theoretical understanding for the observation of 
disordered arrested states not only in colloidal systems 
characterized by a hard core \cite{fabbian,sperl,science,barsh} or 
charge-stabilized
colloidal dispersions \cite{lai:papers}
but also in ultrasoft systems like star polymer solutions. 

We thank H.\ L{\"o}wen for a critical reading of the manuscript.
The Roma group acknowledges support from 
MIUR FIRB and COFIN and INFM PRA HOP and GenFdT. 
C.N.L.\ acknowledges support
by the Deutsche Forschungsgemeinschaft through the SFB TR6.

%
%


\begin{thebibliography}{99}
\bibitem{md} The studied system is composed by
1000 particles of unitary mass $m$ interacting via the 
potential of Eq.\ (\ref{eq-potential}). Distances are measured in
unit of $\sigma$ and  time in units of  $(m \sigma^2 \beta)^{1/2}$,
with $\beta=1$. Integration time step is 0.005.
\bibitem{xtal} The studied ($f,\eta$)  
values correspond to states both in the equilibrium and in the supercooled
region of the phase diagram. We report only 
state points where crystallization did not 
intervene within the time length of the simulation. 
Crystallization has been detected via the 
time evolution of the potential energy and of $S(q)$.
\end{thebibliography}
\end{document}